\documentclass{article}
\usepackage{spconf,amsmath,graphicx,float,multirow}

\title{Weak Alignment Supervision from Hybrid Model \\ Improves End-to-End ASR}
\name{Jintao Jiang$^1$, Yingbo Gao$^2$, Zoltan Tuske$^1$}
\address{$^1$AppTek, McLean, Virginia, USA \\ $^2$AppTek GmbH, 52062 Aachen, Germany \\
\small \texttt{\{jjiang|ygao|ztuske\}@apptek.com}}
\begin{document}
%
\maketitle
\begin{abstract}
In this paper, we aim to create weak alignment supervision from an existing hybrid system to aid the end-to-end modeling of automatic speech recognition. Towards this end, we use the existing hybrid ASR system to produce triphone alignments of the training audios. We then create a cross-entropy loss at a certain layer of the encoder using the derived alignments. In contrast to the general one-hot cross-entropy losses, here we use a cross-entropy loss with a label smoothing parameter to regularize the supervision. As a comparison, we also conduct the experiments with one-hot cross-entropy losses and CTC losses with loss weighting. The results show that placing the weak alignment supervision with the label smoothing parameter of 0.5 at the third encoder layer outperforms the other two approaches and leads to about 5\% relative WER reduction on the TED-LIUM 2 dataset over the baseline. We see similar improvements when applying the method out-of-the-box on a Tagalog end-to-end ASR system.

\end{abstract}
\section{Introduction}
\label{sec:intro}

In recent years, end-to-end (E2E) automatic speech recognition (ASR) modeling has gained tremendous successes \cite{prabhavalkar2023end,li2022recent,bahdanau2016end,lu2016training}. Compared to hybrid modeling, E2E modeling does not need expert language knowledge and the cumbersome HMM training \cite{gales2008application}. The major E2E architectures are Attention-based Encoder Decoder (AED) \cite{chan2016listen,zeyer2019comparison}, Connectionist Temporal Classification (CTC) \cite{graves2006connectionist,graves2014towards,lee2021intermediate,kim2017joint}, and Recurrent Neural Networks-Transducer (RNN-T) \cite{graves2012sequence,rao2017exploring,he2019streaming,liu2021improving,jain2019rnn}. E2E modeling can easily be combined with various algorithms/models/processes, such as attention \cite{vaswani2017attention}, self-supervised learning (SSL) \cite{baevski2020wav2vec,sang2022self}, pretraining \cite{zeyer2019comparison}, labeling smoothing \cite{gao2020towards}, multi-task learning \cite{kim2017joint,boyer2021study,luong2015multi,crawshaw2020multi,zhang2021survey,sogaard2016deep,yang2020predicting}, and data augmentation \cite{sang2022self,cui2015data,park2019specaugment,ruder2017overview}.

Due to the simplicity of E2E modeling, researchers have been moving away from hybrid ASR modeling. However, in cases where a hybrid system is available, why do we throw away the hybrid ASR model instead of making use of it for the E2E modeling? In this paper, we aim at utilizing the existing hybrid ASR models to aid the E2E modeling \cite{liu2021improving,sogaard2016deep,toshniwal2017multitask}. Towards this end, we use an existing hybrid ASR system to align the training audios to derive the frame-by-frame triphone/senones alignment. Then we use the alignments to create an auxiliary task with frame-wise cross-entropy (CE) loss function. This basic idea has been explored in the past \cite{liu2021improving,sogaard2016deep,toshniwal2017multitask}. However, in this work, we propose a novel approach to combine the multi-task learning and label smoothing \cite{gao2020towards,szegedy2016rethinking}. We think the alignment supervision should not overwhelm the primary ASR objective and a weak supervision should be beneficial. In multi-task learning, usually CTC or one-hot CE losses with loss weighting are used. Here, we propose to use a weak CE loss through adjusting the label smoothing parameter. That is, we use a label smoothing parameter of 0.5 to create a weak supervision that produces robust and consistent Word Error Rate (WER) improvements.

\section{Related Work}
\label{sec:relatedwork}
Multi-task learning has been widely used in E2E modeling \cite{kim2017joint,boyer2021study,luong2015multi,crawshaw2020multi,zhang2021survey,sogaard2016deep,chen2023improving}, and some researchers have tried using the state/triphone/senones alignments from hybrid ASR models for the auxiliary tasks \cite{liu2021improving,sogaard2016deep,toshniwal2017multitask}. 

In \cite{toshniwal2017multitask}, The authors use lower-level tasks, such as state or phoneme recognition, in a multi-task training approach with an AED model for direct character transcription. The AED encoder has four Bidirectional Long Short-Term Memory (BLSTM) layers with time-domain sub-sampling \cite{hochreiter1997long,graves2005bidirectional}. They experiment with different auxiliary tasks: a phoneme decoder loss, a phoneme CTC loss, and a frame-wise state-level CE loss. For the phoneme decoder and phoneme CTC losses, the phoneme sequences are produced by composing the transcripts with the pronunciation dictionary. This approach has a limitation in the sense that there are usually pronunciation variants for many of the English words. For the frame-wise state-level CE loss, the authors obtain state-level alignments from a Kaldi HMM/DNN ASR system. Their results on Switchboard and CallHome show consistent improvements over baseline E2E models, obtaining the best performance with a combination of a phoneme decoder at Layer 3 and a frame-level state loss at Layer 2.

In \cite{liu2021improving}, the authors attempt using two auxiliary tasks to improve their streaming RNN-T models. For one auxiliary task, they use an auxiliary RNN-T task at an intermediate encoder layer and exploit a symmetric Kullback–Leibler divergence loss between the output posterior distributions of primary and auxiliary tasks. Similar to \cite{toshniwal2017multitask}, the authors also use state/triphone/senones alignments derived from existing hybrid ASR models. The RNN-T encoder has five latency-controlled BLSTM layers with time-domain sub-sampling after the first and second layers. The primary task uses wordpieces as targets. The main difference from \cite{toshniwal2017multitask} is that \cite{liu2021improving} has a linear+rectified layer before the CE loss, tunes the weights of the auxiliary tasks, and applies them at the third and fifth layers. \cite{liu2021improving} shows that the best results are obtained with loss weighting of 0.6.

The paper \cite{zeyer2019comparison} is of great interest here as its authors share a full recipe for replicating their AED modeling experiments and their TensorFlow based deep learning platform RETURNN \cite{zeyer2018returnn} at GitHub. Besides, the paper describes all the components needed for an E2E ASR modeling: AED, RNN-based decoder, dropout \cite{srivastava2014dropout}, time-domain sub-sampling, Byte Pair Encoding (BPE) \cite{sennrich2015neural}, CTC-based multi-task learning, label smoothing, pretraining, feedback/fertility modeling, learning rate control, BLSTM/Transformer based encoder layers, SpecAugment, Scheduled Sampling \cite{bengio2015scheduled}, etc. 

\section{Methodology}
\label{sec:methodology}\
Auxiliary supervision with loss weighting has been widely used in the literature \cite{lee2021intermediate,kim2017joint,liu2021improving,crawshaw2020multi,toshniwal2017multitask} to regularize the strength of the auxiliary tasks. 

\begin{equation}
   L_\text{all} = L_\text{ASR} + \lambda L_\text{AUX}
   \label{eq: gradientweighting}
\end{equation}

Here, we propose to use the label smoothing to modulate the strength of the auxiliary tasks.

\begin{equation}
   L_\text{all} = L_\text{ASR} + L_\text{AUX}(m_\text{labelsmoothing})
   \label{eq: labelsmoothing}
\end{equation}

For each frame, the auxiliary loss with label smoothing can be further expressed as \cite{vaswani2017attention,gao2020towards}:
\begin{equation}
  \begin{split}
    & L_\text{AUX}(m) = \\
    & -\sum_{v=1}^V{\left((1-m)p_{v}+\frac{m}{V-1}(1-p_{v})\right)\log{q_{v}}} \\
    &= (1-m)L_{\text{CE}} + m L_{\text{smooth}}
    \label{eq: labelsmoothingequation}
  \end{split}
\end{equation}

\noindent where v is a running index in the target vocabulary V, $m$ is the hyper-parameter (between 0 and 1) that controls the amount of probability mass to discount, $p_{v}$ is the one-hot true target distribution and $q_{v}$ is the output distribution of the model.

That is, instead of explicitly weighting the auxiliary losses using $\lambda$ in Equation \ref{eq: gradientweighting}, we implicitly adjust the auxiliary loss strengths through label smoothing using $m$ in Equations \ref{eq: labelsmoothing} and \ref{eq: labelsmoothingequation}, with the hypothesis that higher label smoothing parameters produce weaker supervision \cite{gao2020towards}. Interestingly, the label smoothing loss can be considered as a weighted combination of the one-hot CE loss ($L_{\text{CE}}$) and the uniform CE loss ($L_{\text{smooth}}$). Here, the latter refers to that the target token's true probability is set to 0, while non-target tokens equally share the probability of 1. Therefore, we can use the label smoothing parameter $m$ to control the balance between the discrimination loss ($L_{\text{CE}}$) and the smoothing loss ($L_{\text{smooth}}$).

\begin{figure}[h!]
    \centering
    \includegraphics[scale=0.39]{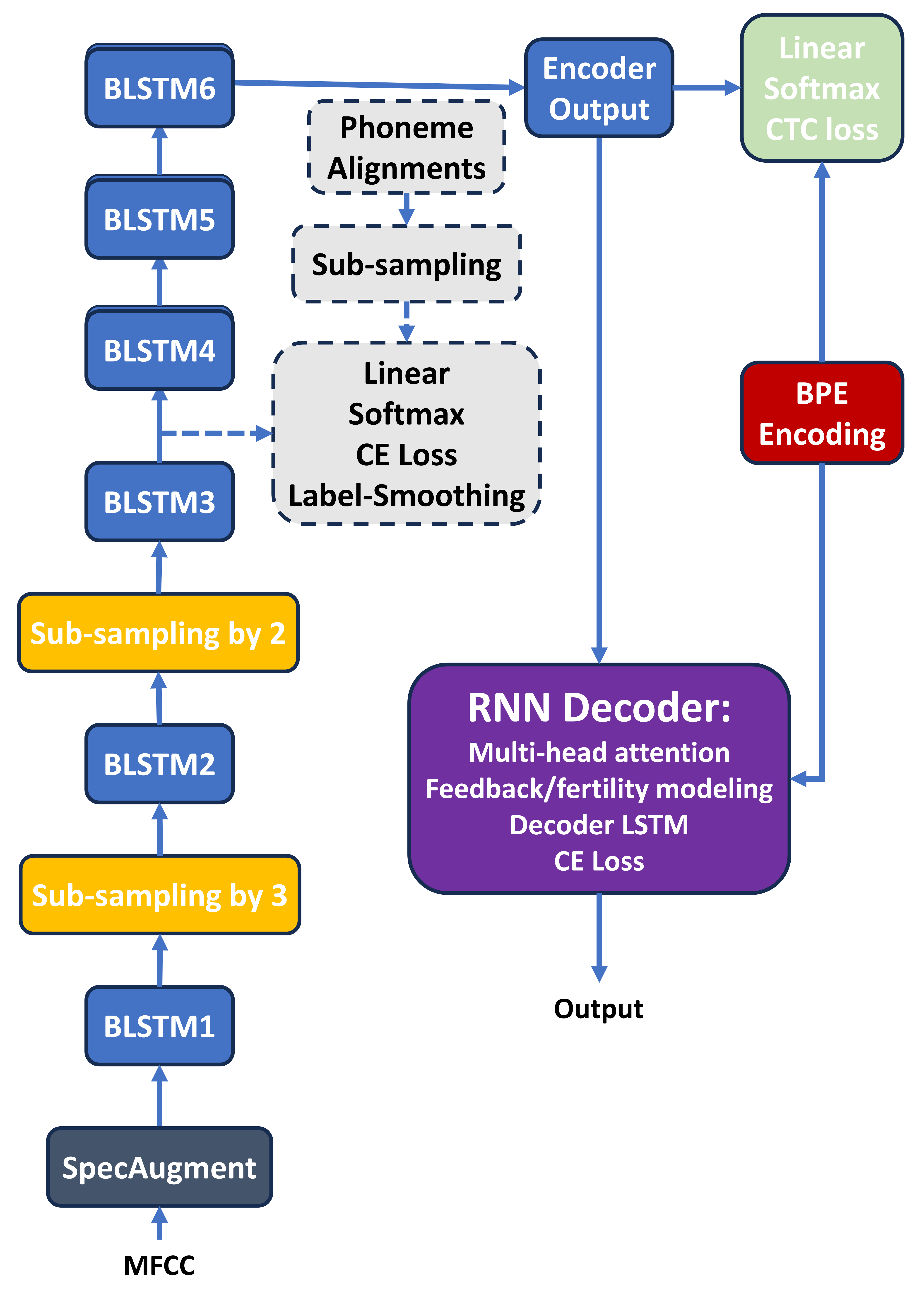}
    \caption{Diagram of the network.}
    \label{fig:diagram}
\end{figure}

We use the recipe from \cite{zeyer2019comparison} with a BLSTM encoder as our baseline AED ASR system.
Figure \ref{fig:diagram} shows the diagram of the original networks and the proposed frame-wise CE loss with the phoneme alignments. 

MFCC features extracted from audio files are sent to SpecAugment \cite{park2019specaugment} and then to the six BLSTM layers. At the first and second layers, time-domain sub-sampling is applied. The output from the sixth BLSTM goes to a linear encoder that connects to an existing auxiliary CTC loss with BPEs as targets, and the encoder output also connects to the RNN decoder. Note that the CTC loss output is not used in decoding. Inside the decoder, there are multi-head attention computation, feedback/fertility modeling, decoder LSTM, and a CE loss with the BPEs as targets. The dotted lines and boxes indicate the proposed CE loss with triphone alignments and label smoothing. The proposed auxiliary task can be put after BLSTM Layers 1, 2, 3, 4, 5, and the encoder output layer. For the label smoothing parameter, we test different values of 0.0, 0.1, 0.3, 0.5, 0.7, 0.9, and 1.0. As in \cite{zeyer2019comparison}, pretraining is used to help convergence. As a comparison, the CE loss with triphone alignments and loss weighting is applied at different layers and with different weights of 0.05, 0.1, 0.2, 0.5, and 1.0. To examine whether there is any benefit of introducing triphone alignments, we also experiment with the usual auxiliary CTC loss task with loss weighting.

\section{Experimental Results}
\label{sec:results}

\subsection{Setup}
\label{ssec:setup}
The experiments are performed on the TED-LIUM release 2 (200 hours) \cite{rousseau2014enhancing} using RETURNN \cite{zeyer2018returnn}. We follow a recipe from \cite{zeyer2019comparison} \footnote{https://github.com/rwth-i6/returnn-experiments/tree/master/2019-asr-e2e-trafo-vs-lstm/tedlium2} with a specific config file \footnote{base2.specaug.bs18k.curric3.pfixedlr.lrd07.eos.config}. In the repository, a full setup is provided. For this dataset, we use the same train/dev/test division as in \cite{zeyer2019comparison}. The RETURNN platform builds on TensorFlow \cite{abadi2016tensorflow}. Note that in this study, language models are not applied. We acknowledge that stronger baselines are possible, e.g. with Conformer/E-Branchformer \cite{kim2023branchformer,zeineldeen2023chunked} and wave2vec 2.0 \cite{baevski2020wav2vec}, however, we think this baseline is sufficient for the purpose of this work.

As in \cite{zeyer2019comparison}, we limit the training time by a fixed number of epochs (37.5) over the training dataset in all experiments. (There are 150 sub-epochs in total.) The training takes about 32 hours on a single NVIDIA 1080 Ti GPU.

MFCC features are extracted with a 25-ms window, a 10-ms step size, and 40 Mel-filters. Before training, the mean and standard deviation statistics for training data are estimated and thus used in pre-processing for feature normalization.

Triphone alignments are obtained using an existing hybrid BLSTM ASR model that uses 5000 triphones, 80-dimension MFCC as acoustic features, and four 1024-dimension hidden BLSTM layers. The hybrid model was trained using about 11,000 hours of training data. The TED-LIUM release 2 data were included in the training. Traditional Gaussian Mixture Models (GMM) was trained to obtain the triphone CART tree and then the BLSTM model was trained with SpecAugment using RETURNN \cite{zeyer2018returnn}. The RASR toolkit \cite{rybach2011rasr} is used to perform the forward-backward alignments.

As shown in Figure \ref{fig:diagram}, the BLSTM layers have a hidden dimension of 1024. We use dropout and Adam optimizer \cite{kingma2014adam} with a variant of Newbob learning rate scheduling. In the RNN decoder, the CE loss is applied with a label smoothing of 0.1. During decoding, a beam size of 12 is used. WER is computed for the decoder output (after merging BPEs into words) using NIST SCLITE Scoring Toolkit \cite{sclite}.

\subsection{Triphone CE losses with loss weighting}
\label{ssec:CEscale}

\begin{figure}[H]
    \centering
    \includegraphics[scale=0.8]{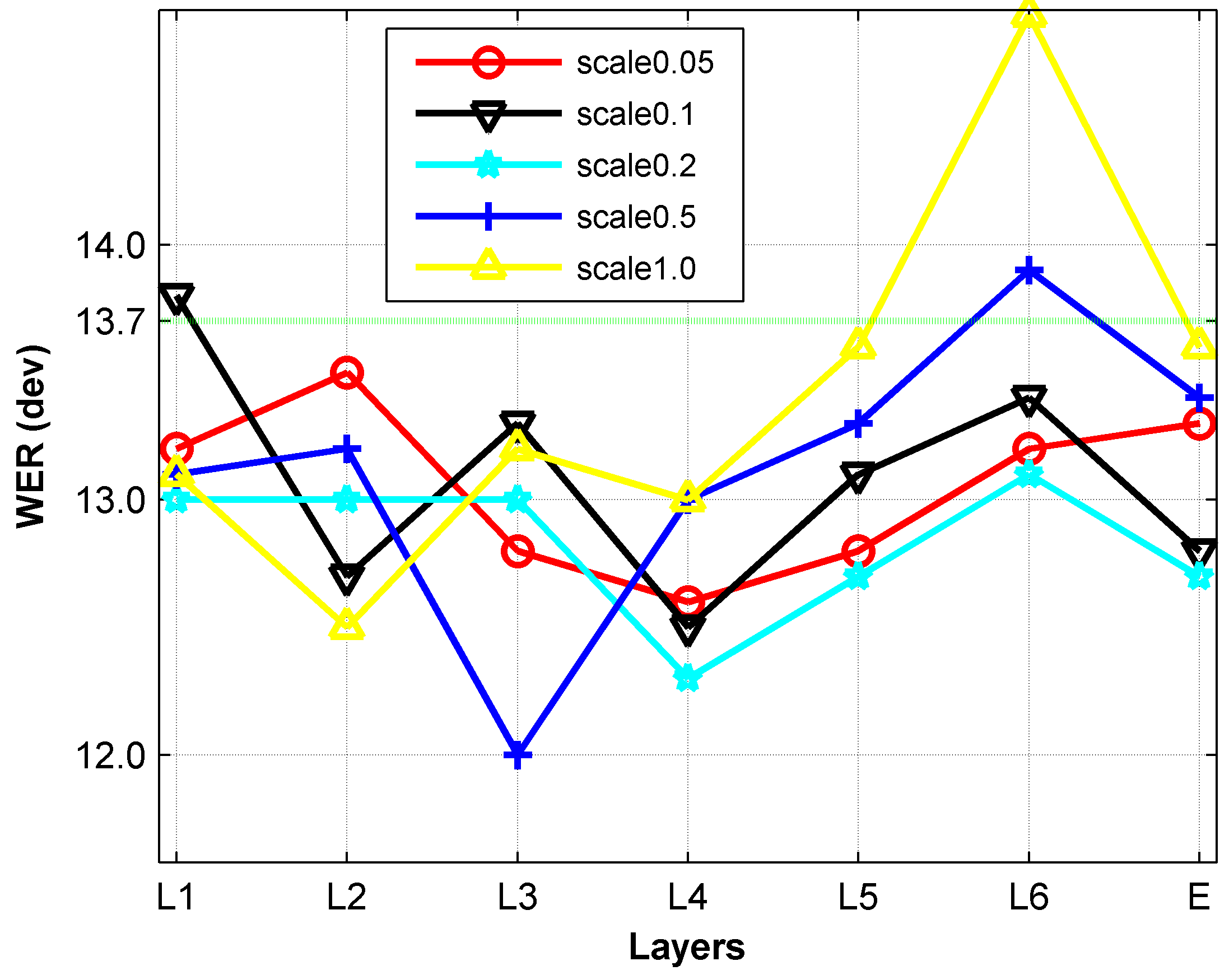}
    \caption{WERs for the dev dataset with different loss weights and at different encoder layers. L1, L2, L3, L3, L4, L5, and L6 refer to the BLSTM layers 1, 2, 3, 4, 5, and 6, respectively. E refers to the encoder output layer. The green horizontal line indicates the baseline.}
    \label{fig:dev_wer_scale}
\end{figure}

We first apply the triphone CE losses with loss weighting (0.05, 0.1, 0.2, 0.5, and 1.0) at different encoder BLSTM layers. Figure \ref{fig:dev_wer_scale} shows WERs for the dev dataset. When the loss weighting is 1.0, the best result is obtained when placing the auxiliary task at the second BLSTM layer that is consistent with results from \cite{toshniwal2017multitask}. When the loss weighting is 0.5, the best result is obtained when placing the auxiliary task at the third BLSTM layer. Subsequently, when the loss weights are 0.2, 0.1, and 0.05, the auxiliary task at the fourth layer produces the best results. It is interesting to notice the trend here that when the loss weighting is higher, the auxiliary task should be placed at the lower encoder layers (e.g., the second), and when the loss weighting is lower, the auxiliary task should be placed at the higher encoder layers (e.g., the fourth). Overall, a loss weighting of 0.5 at the third encoder BLSTM layer gives the best results, and this is again consistent with \cite{liu2021improving} that uses a CE loss task with loss weighting of 0.6 at the third and fifth layers. Therefore, given that triphone target in the auxiliary task is different from the BPE target in the primary task, the auxiliary task requires a weak supervision that can be achieved either through loss weighting, being far away from the primary task, or using the proposed label smoothing.

\begin{figure}[H]
    \centering
    \includegraphics[scale=0.8]{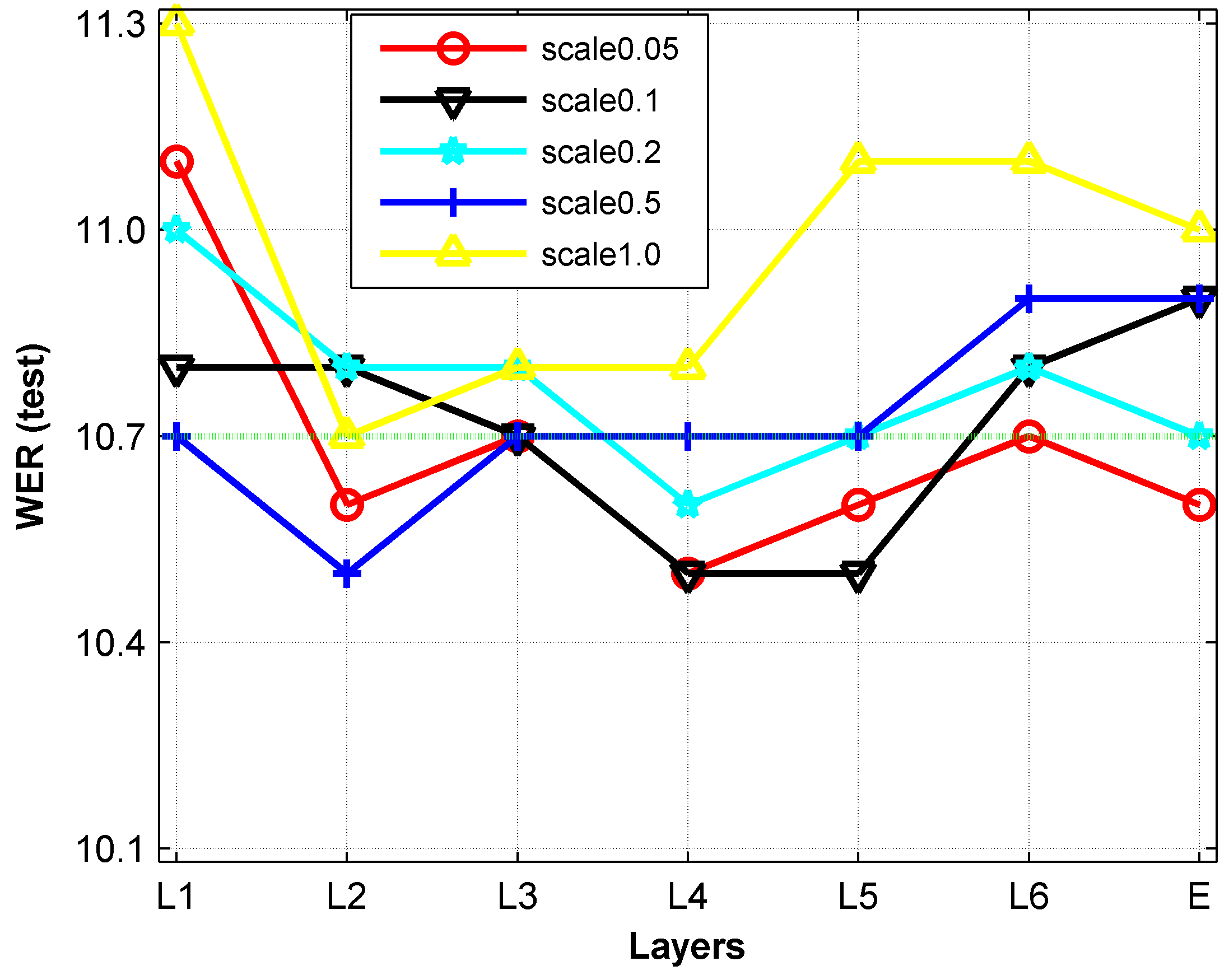}
    \caption{WERs for the test dataset. Note that this figure is shown for information completeness only and the loss weighting and layer selection are not based on it.}
    \label{fig:eval_wer_scale}
\end{figure}

We run the same tests on the test dataset and the results are shown in Figure \ref{fig:eval_wer_scale}. When the loss weighting is 1.0, the best result is obtained when placing the triphone CE loss at the second layer that is consistent with results for the dev dataset. However, for the test dataset, the triphone CE loss with a loss weighting of 1.0 at the second layer does not improve over the baseline. Similarly, the loss weighting of 0.5 CE loss at the third layer does not improve over the baseline for the test dataset. However, with the loss weights of 0.05 and 0.2, the auxiliary tasks at the fourth layer produce the best results (better than the baseline) for the test data. For these specific cases (loss weights of 0.05 and 0.2), the results for the dev dataset are consistent with those for the dev dataset. The results are summarized in Table \ref{tab:scale}. Overall, the performance patterns from the loss weighting and triphone auxiliary task placement interaction differ across the dev and test datasets, indicating that the loss weighting is not a robust approach to regularize the triphone CE losses. 

\begin{table}[H]
\center
\begin{tabular}{|c|c|c|c|}
\hline
\multirow{2}{*}{\begin{tabular}[c]{@{}c@{}}Loss\\ Weighting\end{tabular}} & \multirow{2}{*}{\begin{tabular}[c]{@{}c@{}}Best CE layer\\ (based on dev)\end{tabular}} & \multicolumn{2}{c|}{WER} \\ \cline{3-4}
& & dev & test \\ \hline \hline
Baseline & - & 13.7 & 10.7 \\ \hline
0.05 & 4 & 12.6 & 10.5 \\ \hline
0.1 & 4 & 12.5 & 10.5 \\ \hline
0.2 & 4 & 12.3 & 10.6 \\ \hline
0.5 & 3 & 12.0 & 10.7 \\ \hline
1.0 & 2 & 12.5 & 10.7 \\ \hline
\end{tabular}
\caption{CE loss weighting results}
\label{tab:scale}
\end{table}

\subsection{CTC losses with loss weighting}
\label{ssec:CTCloss}
\begin{figure}[H]
    \centering
    \includegraphics[scale=0.8]{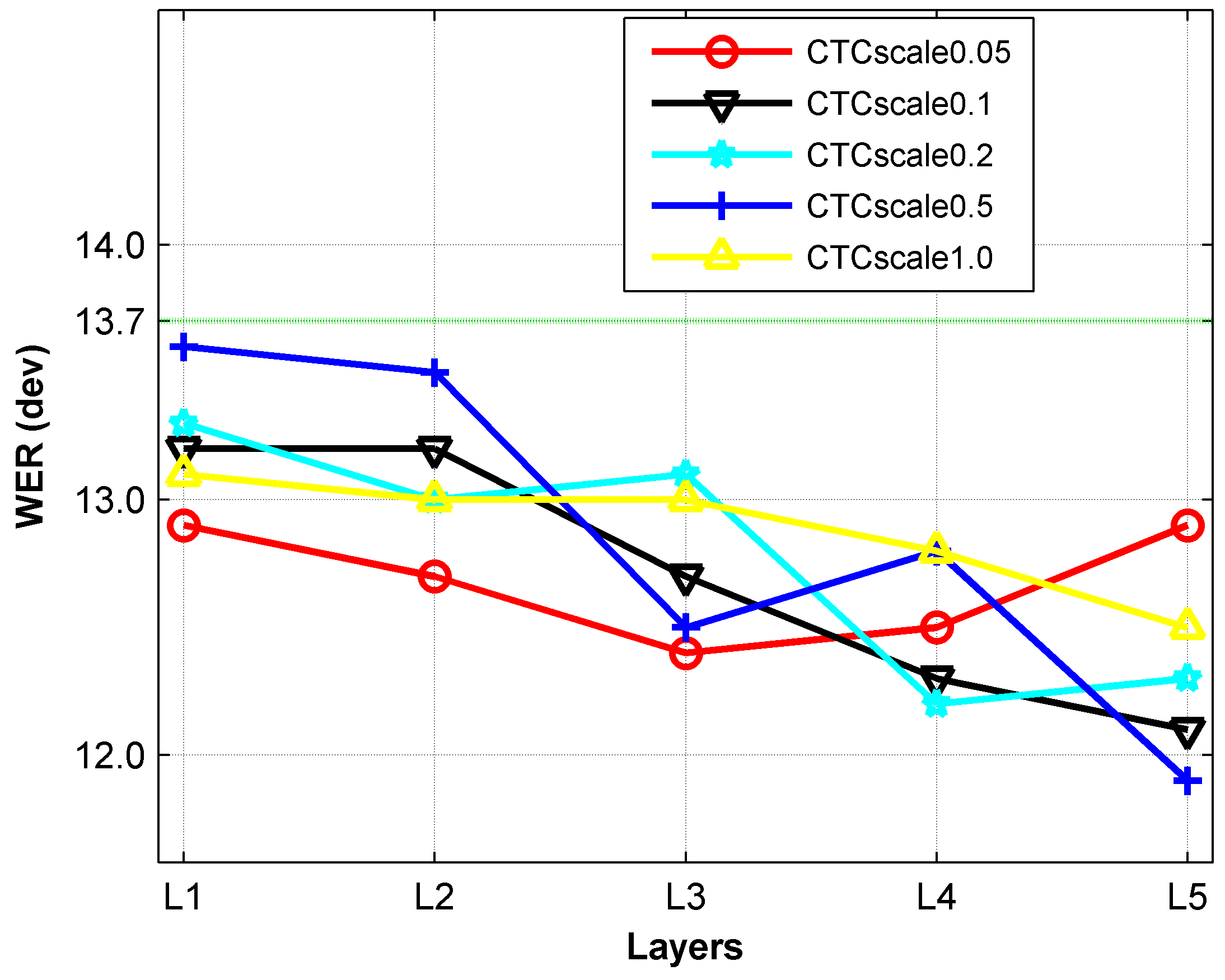}
    \caption{WERs for the dev dataset with different CTC loss weights and at different encoder layers.}
    \label{fig:dev_wer_CTCscale}
\end{figure}

\begin{figure}[H]
    \centering
    \includegraphics[scale=0.8]{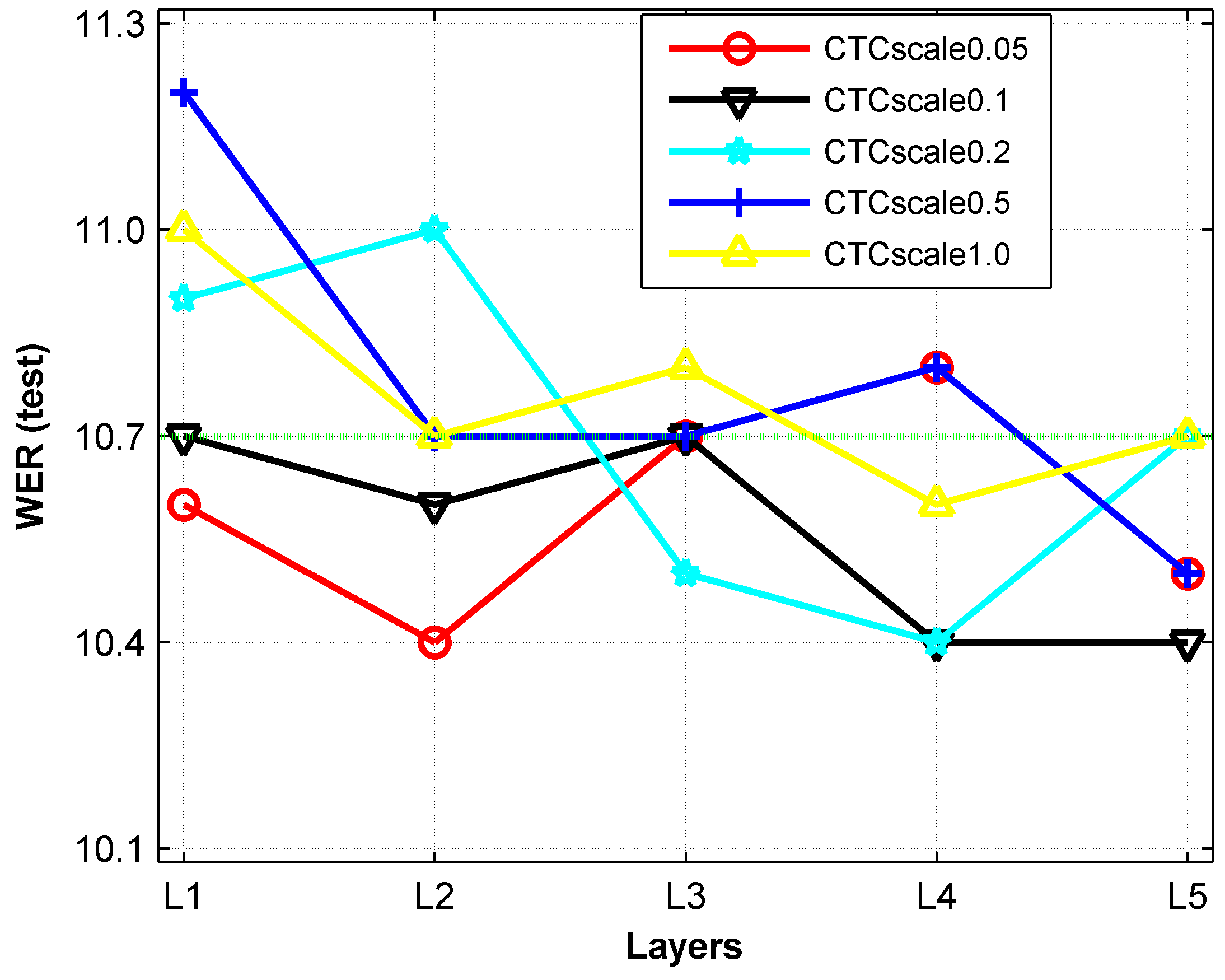}
    \caption{WERs for the test dataset.}
    \label{fig:eval_wer_CTCscale}
\end{figure}

We use the BPEs that are already prepared for the primary task to create the proposed auxiliary CTC task. Note that in Figure \ref{fig:diagram}, an auxiliary CTC loss without loss weighting at the encoder output layer exists already. Therefore, when testing auxiliary CTC tasks, we only place it at the encoder BLSTM layers 1-5, and not on top of the whole encoder stack. The decoding results are shown in Figures \ref{fig:dev_wer_CTCscale} and \ref{fig:eval_wer_CTCscale}.

Figure \ref{fig:dev_wer_CTCscale} shows that overall CTC losses should be placed at the higher encoder BLSTM layers, compared to those in Figure \ref{fig:dev_wer_scale}. Specifically, when the loss weights are 0.05, 0.1, 0.2, 0.5, and 1.0, the corresponding best layer placement are 3, 5, 4, 5, and 5, respectively. 

\begin{table}[H]
\center
\begin{tabular}{|c|c|c|c|}
\hline
\multirow{2}{*}{\begin{tabular}[c]{@{}c@{}}Loss\\ Weighting\end{tabular}} & \multirow{2}{*}{\begin{tabular}[c]{@{}c@{}}Best CTC layer\\ (based on dev)\end{tabular}} & \multicolumn{2}{c|}{WER} \\ \cline{3-4}
& & dev & test \\ \hline \hline
Baseline & - & 13.7 & 10.7 \\ \hline
0.05 & 3 & 12.4 & 10.7 \\ \hline
0.1 & 5 & 12.1 & 10.4 \\ \hline
0.2 & 4 & 12.2 & 10.4 \\ \hline
0.5 & 5 & 11.9 & 10.5 \\ \hline
1.0 & 5 & 12.5 & 10.7 \\ \hline
\end{tabular}
\caption{CTC loss weighting results}
\label{tab:CTCscale}
\end{table}

Similarly to Section \ref{ssec:CEscale}, we create a table to summarize the results from Figures \ref{fig:dev_wer_CTCscale} and \ref{fig:eval_wer_CTCscale}. Generally, the auxiliary CTC loss should be placed at higher layers. Again, this is consistent with the observations in \cite{toshniwal2017multitask}. For lower loss weights, the CTC loss should be placed at the lower BLSTM layers. This pattern is different from that of CE loss in Section \ref{ssec:CEscale}. Overall, the performance patterns from the loss weighting and the CTC auxiliary task placement interaction differ across the dev and test datasets, indicating that the loss weighting is not a robust approach to regularize the BPE CTC losses.
 
\begin{figure}[H]
    \centering
    \includegraphics[scale=0.8]{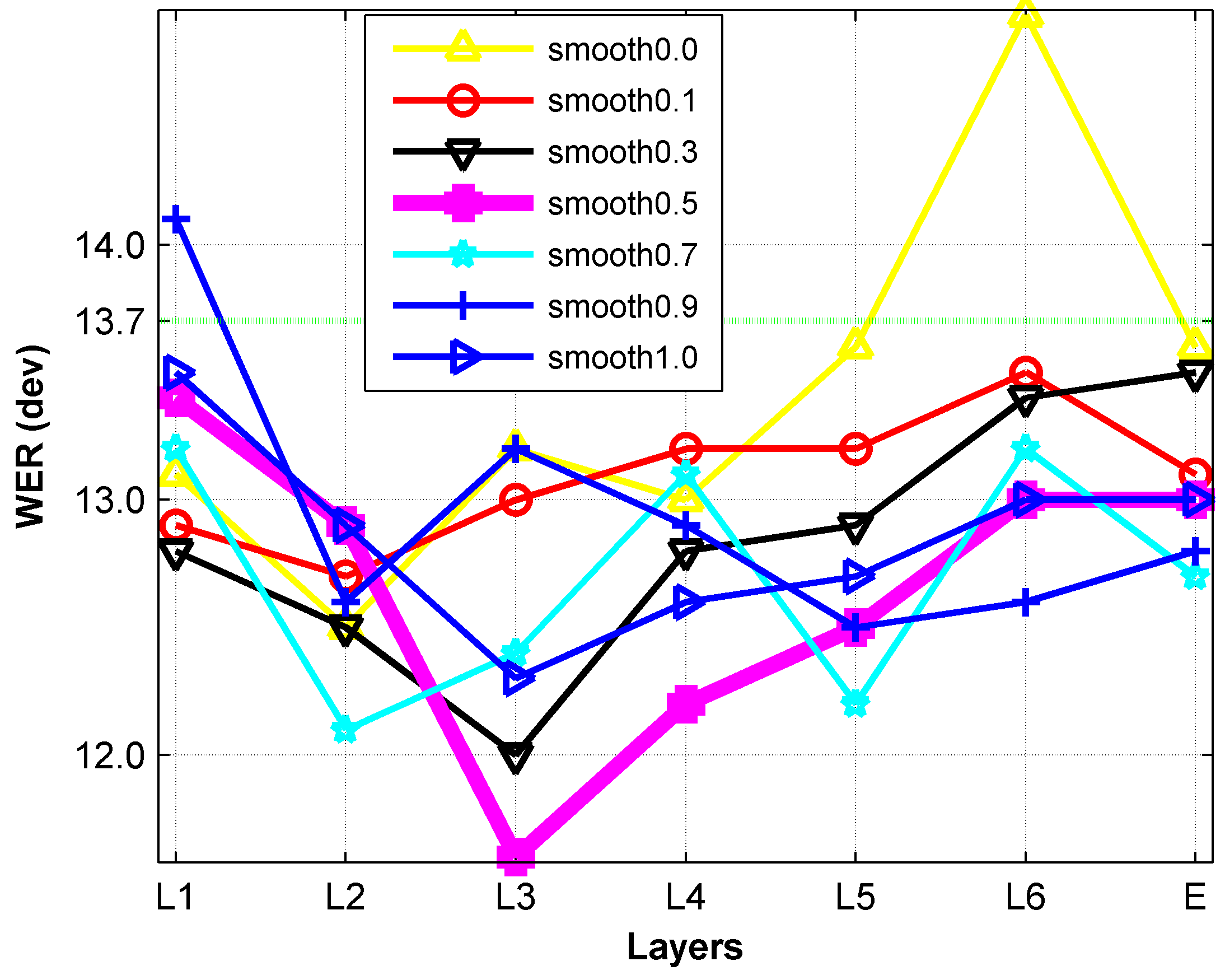}
    \caption{WERs (dev) with different $m$'s and at different layers.}
    \label{fig:dev_wer_smooth}
\end{figure}

\begin{figure}[H]
    \centering
    \includegraphics[scale=0.8]{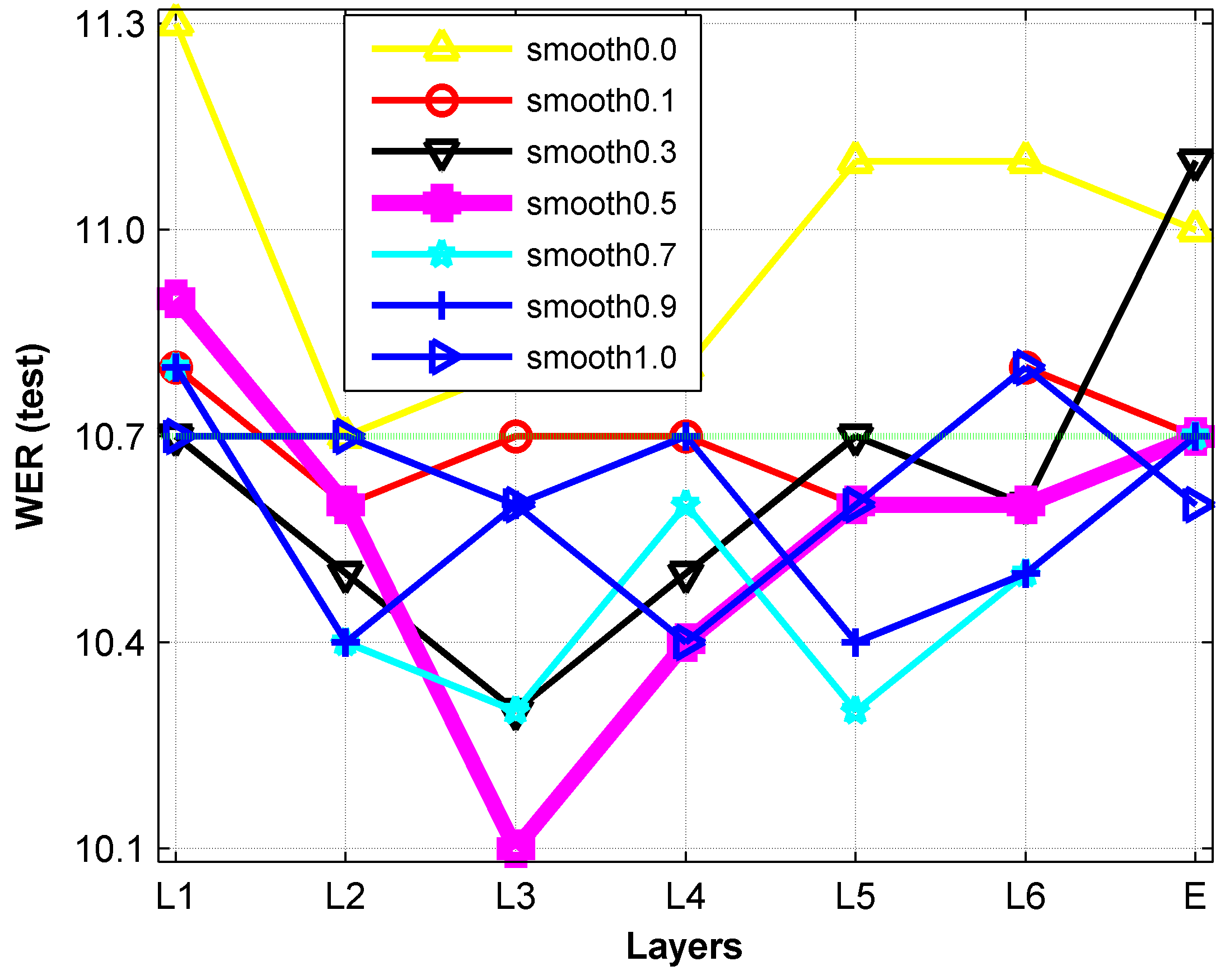}
    \caption{WERs (test) with different $m$'s and at different layers.}
    \label{fig:eval_wer_smooth}
\end{figure}

\subsection{Triphone CE losses with label smoothing}
\label{ssec:smooth}

In this experiment, we focus on the auxiliary CE loss at different encoder layers with label smoothing. Here, we set the loss weighting at 1.0 given that the label smoothing can be considered as another way to regularize the strength of the auxiliary tasks \cite{gao2020towards}. We run the experiments with label smoothing values of 0.0, 0.1, 0.3, 0.5, 0.7, 0.9, and 1.0. Results are shown in Figures \ref{fig:dev_wer_smooth} and \ref{fig:eval_wer_smooth}.

Figure \ref{fig:dev_wer_smooth} shows the WERs obtained for the dev dataset with different label smoothing parameters at different encoder layers. Comparing to Figures \ref{fig:dev_wer_scale} and \ref{fig:dev_wer_CTCscale}, we have the following observations: (1) Label smoothing performs better than loss weighting in terms of regularizing the auxiliary task losses; (2) Label smoothing tends to perform the best in the middle layers (i.e., Layer 3 here); (3) There exists an optimal value of the label smoothing parameter, which is \textbf{0.5}; and (4) Label smoothing produces similar results across the dev and test datasets, indicating that label smoothing is a robust and effective way of regularizing auxiliary task losses. This supports our thinking that a relatively weak and robust supervision should be preferred when augmenting the training with auxiliary tasks. At the same time, the proposed approach is different from others \cite{toshniwal2017multitask,liu2021improving} in the sense that it directly adjusts the targets' true probability to 0.5, a value that is optimal in balancing the discriminative loss ($L_{\text{CE}}$) and the smoothing loss ($L_{\text{smooth}}$) (see Equation \ref{eq: labelsmoothingequation}).

\begin{table}[H]
\center
\begin{tabular}{|c|c|c|c|}
\hline
\multirow{2}{*}{\begin{tabular}[c]{@{}c@{}}Label\\ Smoothing\end{tabular}} & \multirow{2}{*}{\begin{tabular}[c]{@{}c@{}}Best CE layer\\ (based on dev)\end{tabular}} & \multicolumn{2}{c|}{WER} \\ \cline{3-4}
& & dev & test \\ \hline \hline
Baseline & - & 13.7 & 10.7 \\ \hline
0.0 & 2 & 12.5 & 10.7 \\ \hline
0.1 & 2 & 12.7 & 10.6 \\ \hline
0.3 & 3 & 12.0 & 10.3 \\ \hline
\textbf{0.5} & \textbf{3} & \textbf{11.6} & \textbf{10.1} \\ \hline
0.7 & 2 & 12.1 & 10.4 \\ \hline
0.9 & 5 & 12.5 & 10.4 \\ \hline
1.0 & 3 & 12.3 & 10.6 \\ \hline
\end{tabular}
\caption{Label smoothing results}
\label{tab:smooth}
\end{table}

\begin{table}[H]
\center
\begin{tabular}{|c|c|c|c|}
\hline
\multirow{2}{*}{\begin{tabular}[c]{@{}c@{}}Loss\\ Regularization\end{tabular}} & \multirow{2}{*}{\begin{tabular}[c]{@{}c@{}}Best paramemeters\\ (based on dev)\end{tabular}} & \multicolumn{2}{c|}{WER} \\ \cline{3-4}
& & dev & test \\ \hline \hline
Label Smoothing & \textbf{$m$=0.5, layer=3} & \textbf{11.6} & \textbf{10.1} \\ \hline
CE LW & weight=0.5, layer=3 & 12.0 & 10.7 \\ \hline
CTC LW & weight=0.5, layer=5 & 11.9 & 10.5 \\ \hline
\end{tabular}
\caption{System comparison (LW = Loss Weighting)}
\label{tab:comparison}
\end{table}

We summarize the results from Figures \ref{fig:dev_wer_smooth} and \ref{fig:eval_wer_smooth} into Table \ref{tab:smooth}. Note the label smoothing parameter of 0.0 is equal to the CE loss with a loss weight of 1.0 which serves as a sanity check. That is, the yellow lines in Figures 6 and 7 correspond to those in Figures 2 and 3, respectively. For the label smoothing parameter of 1.0, according to Equation \ref{eq: labelsmoothingequation}, it produces a minimally discriminative loss objective where the smoothing loss ($L_{smooth}$) overtakes the CE loss ($L_{CE}$). Nevertheless, such non-discriminative loss objective still produces small improvement over baseline. Table \ref{tab:smooth} clearly shows that the label smoothing parameter of 0.5 produces the overall best results at Layer 3 with a relative 5.6\% in WER reduction on the test dataset. Interestingly, the label smoothing parameters of 0.3 and 0.7, the left and right sides of neighbors of the optimal parameter of 0.5, produce the second-to-the-best results at Layer 2.

Based on the results on the dev dataset, we select the best setups from Tables \ref{tab:scale}, \ref{tab:CTCscale}, and \ref{tab:smooth} and put the corresponding results in Table \ref{tab:comparison}. Obviously, the label smoothing approach of loss regularization outperforms the other two. 

\subsection{Applying to the Tagalog ASR}

\begin{table}[H]
\center
\begin{tabular}{|c|c|c|}
\hline
\multirow{2}{*}{\begin{tabular}[c]{@{}c@{}}Systems\end{tabular}} & \multicolumn{2}{c|}{WER} \\ \cline{2-3}
& dev & test \\ \hline \hline
Baseline & 20.6 & 25.1 \\ \hline
Auxiliary CE loss (smooth=0.5) & 19.9 & 23.9 \\ \hline
Relative improvement & 3.4\% & 4.8\% \\ \hline
\end{tabular}
\caption{WERs for Tagalog ASR}
\label{tab:tagalog}
\end{table}

We take the best configuration from Section \ref{ssec:smooth}, that is, an auxiliary CE loss with label smoothing parameter of 0.5 at Layer 3, to our E2E Tagalog ASR modeling. The data are divided into three subsets: train (232 hours), dev (22 hours), and test (11 hours). Our existing hybrid Tagalog ASR model was trained with the same datasets. The hybrid model was trained with the same network architecture as in Section \ref{ssec:setup} except that 4500 triphones were used here. Then, we use the existing hybrid Tagalog ASR to align the training audios to produce triphone alignments. We compute 1K BPE from the training transcripts. After the E2E training, the recognition results are presented in Table \ref{tab:tagalog}. For the test dataset, the proposed approach leads to 4.8\% relatively in WER reduction, confirming that the proposed weak triphone alignment supervision is robust and effective in improving E2E ASR modeling.

\section{Conclusion and Discussion}
\label{sec:conclusion}
In this work, we propose to use label smoothing to construct a weak alignment supervision to aid the E2E ASR modeling. We compare the proposed approach with the CE loss with loss weighting \cite{toshniwal2017multitask,liu2021improving} and CTC loss with loss weighting \cite{liu2021improving}. Our results show that the proposed weak alignment supervision with a label smoothing of 0.5 produces consistent and robust results and outperforms the other two approaches. When the proposed approach with the optimal configuration is applied to model a Tagalog E2E ASR system, we observe similar improvements in WER reduction. Thus, the contribution from this study is to provide a different and novel perspective in designing weak alignment supervision. Future work includes improving the baseline, applying the approach to large datasets, and exploring with Conformer/Ebranchformer network architectures \cite{Gulati2020,kim2023branchformer,majumdar2021citrinet,zeineldeen2023chunked}.


\newcommand{\BIBdecl}{\setlength{\itemsep}{0.05 em}}
\bibliographystyle{IEEEtran}
\bibliography{strings,refs}

\end{document}